\documentclass[sigconf, 10pt]{acmart}
\setcopyright{none}
\renewcommand\footnotetextcopyrightpermission[1]{}  
\AtBeginDocument{%
  }

\usepackage{graphicx}
\usepackage{hyperref}
\usepackage{listings}
\usepackage{xcolor}
\usepackage{tikz}
\usetikzlibrary{arrows.meta, positioning}
\newtheorem{theorem}{Theorem}[section]

\acmConference[arXiv]{arXiv}{May 2025}{arXiv}

\newif\ifnotes
\notestrue

\title[Hypergraph Ranking in Networks]{A Ranking Framework for Network Resource Allocation and Scheduling via Hypergraphs}

\author{Rajpreet Singh$^{1}$, Novak Boškov$^{2}$, Aditya Gudal$^{2}$, Manzoor A. Khan$^{2}$}
\affiliation{%
  \institution{$^1$Technical University of Munich \quad $^2$Nokia Bell Labs}
  \city{}
  \country{}
}
\email{rajpreet.singh@tum.de}
\email{{novak.boskov, aditya.gudal, manzoor.a.khan}@nokia-bell-labs.com}

\renewcommand{\shortauthors}{Rajpreet Singh et al.}

\begin{abstract}
    Resource allocation and scheduling are a common problem in various distributed systems. Although widely studied, the state-of-the-art solutions either do not scale or lack the expressive power to capture the most complex instances of the problem. To that end, we present a mathematical framework for hypergraph ranking and analysis, unifying graph theory, lattice theory, and semantic analysis. In our fundamental theorem, we prove the existence of partial order on entities of hypergraphs, extending traditional hypergraph analysis by introducing semantic operators that capture relationships between vertices and hyperedges. Within the boundaries of our framework, we introduce an algorithm to rank the node-hyperedge pairs with respect to the captured semantics. The strength of our approach lies in its applicability to complex ranking problems that can be modeled as hypergraphs, including network resource allocation, task scheduling, and table selection in Text-to-SQL. Through simulations, we demonstrate that our framework delivers nearly optimal problem solutions at a superior run time performance.
\end{abstract}

\begin{document}

\maketitle






\section{Introduction}
\label{sec:introduction}

Modeling and analyzing complex entity relationships is a foundational challenge in networking, computational linguistics, and knowledge representation~\cite{wang2021ratsqlrelationawareschemaencoding}. Traditional graph-based representations such as the one of Edmonds~\cite{Edmonds_1965}, while powerful, often fail to capture the full complexity of multidimensional interactions inherent in real-world systems. These limitations necessitate innovative frameworks that combine mathematical rigor with computational efficiency.

Historically, resource allocation and scheduling problems rely on evolutionary algorithms, graph-based, and integer linear programming. While these methods provide structured ways to allocate resources and schedule tasks, they often fail to capture the multi-dimensional interactions and higher-order dependencies present in real-world systems. As systems grow in complexity, with intricate dependencies and hierarchical constraints, traditional models become inadequate, necessitating more expressive and mathematically robust frameworks.
For instance, linear programming and convex optimization techniques achieve optimal task distribution while minimizing computational overhead, but face scalability issues when handling large, heterogeneous datasets with multi-dimensional dependencies. 
One significant limitation of traditional approaches is their reliance on static optimization models that assume well-defined constraints and complete information without embedding the semantic meaning. On the other hand, AI-driven approaches show promising improvements due to their capability to learn features in complex datasets, but are computationally inefficient and require intensive datasets for training. In real-world scenarios, resource availability and dependencies fluctuate, requiring adaptive models capable of handling uncertainty and hierarchical structures. Pure DAG-based scheduling methods, while effective in enforcing precedence constraints, often fail to capture the fluid nature of resource inter-dependencies and fail to incorporate multi-layered prioritization. 

To address these challenges, we presents a refined, hypergraph based allocation framework that incorporates hierarchical dependencies and precedence constraints using order-theoretic principles. Our main contributions are as follows:
\begin{itemize}
    \item We introduce a generic ranking framework that leverages the interaction of nodes and hyperedges to encode domain-specific semantics, allowing for a more refined relational modeling. 
    \item We establish a connection between hypergraphs and directed acyclic graphs (DAGs), facilitating topological sorting as a concrete ranking extension.
    \item We ensure computational efficiency and scalability through DAG based ranking strategy.
    \item We apply our mathematical framework to a subclass of concrete resource allocation and scheduling problems and demonstrate the superiority of our approach through simulations.
\end{itemize}

The rest of this paper is organized as follows. In Section~\ref{sec:modeling}, we describe the details of our mathematical modeling, explain its benefit, and give our core theorem and approximation bounds. In Section~\ref{sec:perf}, we analyze the performance implications of our modeling. We present concrete resource allocation and scheduling solutions in Section~\ref{sec:simulation}, and conclude with a brief analysis of related literature Section~\ref{sec:related} and final remarks in Section~\ref{sec:conclusion}.

\section{Modeling}\label{sec:modeling}
The problem objective is to allocate top-$k$ resources to tasks while minimizing the total allocation cost. Using hypergraph representation, this problem can be formulated as a ranking problem based on the composition of nodes and hyperedges. The composition is performed by the operators giving semantic meaning to the interaction between nodes and hyperedges. The operator $\otimes$ acts like a bounded linear operator within a finite-dimensional setting of resource-task pairs. The bounded nature ensures that scores are well-defined and score $\Upsilon$($v$ $\otimes$ $e$) remains finite. The proposed algorithm effectively performs ranked projections of relevance scores onto the hypergraph edges, analogous to projecting a function in Hilbert space onto a finite subspace of orthonormal basis vectors. The operator specific ranking ensures that the top-$k$ resources selected for a task are not arbitrarily but represent semantically aligned resources. This reduces the discrepancy between the optimal solution and the solution given by our algorithm. The approximation factor for the algorithm proposed here is tighter compared to naive weight-based approaches.

\begin{figure*}    
    \centering
    \includegraphics[width=\textwidth]{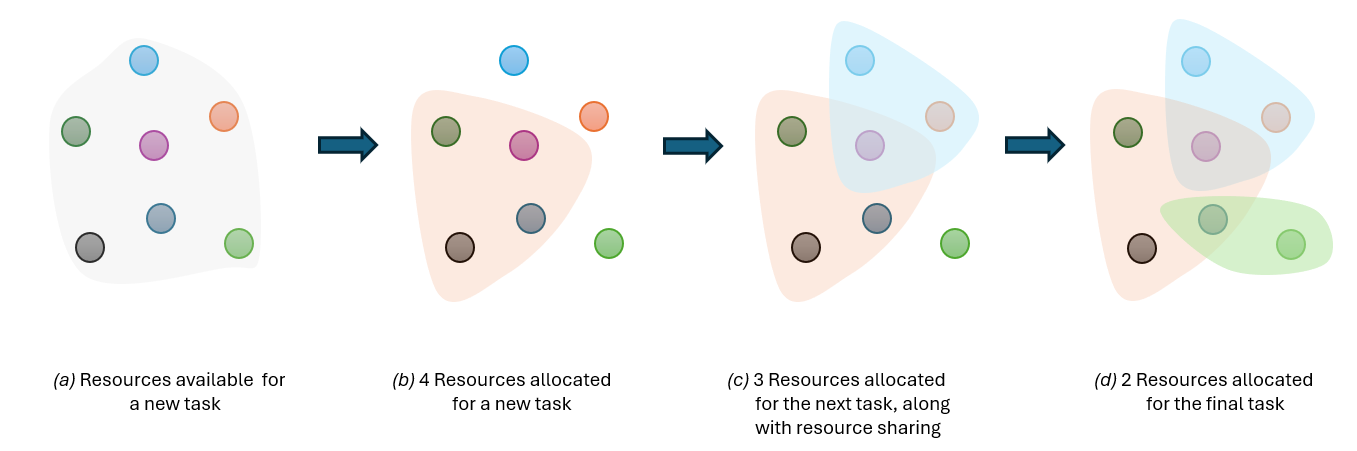}
    \caption{Hypergraph ranking to allocate resources}
    \label{fig:wide1}
\end{figure*}

We can construct the hypergraph-based ranking from a basic resource allocation problem as follows. For the given basic resource allocation where where the given is:
\begin{itemize}
    \item Set of resources: $R = \{1, 2, ..., n\}$
    \item Set of tasks: $T = \{1, 2, ..., m\}$
    \item For each resource $i \in R$:
        \begin{itemize}
            \item $c_i$: cost of resource $i$
            \item $\vec{q_i}$: metadata vector of resource $i$
        \end{itemize}
    \item For each task $j \in T$:
        \begin{itemize}
            \item $k_j$: number of required resources for task $j$
        \end{itemize}
\end{itemize}
Its hypergraph representation is given as:
\begin{itemize}
    \item A hypergraph $H = (V, E)$ where:
        \begin{itemize}
            \item $V$: set of resources
            \item $E$: set of hyperedges representing tasks
        \end{itemize}
    \item For each resource $v_i \in V$:
        \begin{itemize}
            \item $w(v_i)$: cost of resource $i$
            \item $\vec{q_i}$: metadata vector of resource $i$
            \item $\Upsilon$ ($v$, $e$): relevance score of resource $v$ for a task $e$ calculated from ($v$ $\otimes$ $e$)
        \end{itemize}
    \item For each hyperedge $e \in E$:
        \begin{itemize}
            \item $k$: number of required resources
        \end{itemize}
\end{itemize}

The objective function minimizes the total cost:
\[
\min \sum_{i=1}^k  w(v_i),
\]
where $k$ is the number of top ranked resources for a task represented by a hyperedge $e$. The resource selection then solves the following problem. For each task $j$, select $k$ resources with the highest relevance scores (as illustrated in the Fig.~\ref{fig:wide1}):
\[
R_j^* = \arg\max_{S \subseteq V, |S|=k} \sum_{i \in S} v_i.
\]

\subsection{Model Properties \& Benefits}
The hypergraph-based ranking framework is better than traditional graph models because it encodes domain-specific semantics through nodes and hyperedges, allowing for high-dimension relational dynamics that standard edge-based methods fail to capture. By leveraging partial orders and the existence of supremum and infimum, it enables hierarchical and algebraic reasoning, offering a deeper structural understanding beyond simple one-edge connectivity. Unlike conventional ranking models that rely solely on graph-theoretic measures, this approach maps naturally to directed acyclic graphs (DAGs), allowing topological sorting for computationally efficient realizations of ranking extensions. It also ensures a rigorous yet flexible representation of constraints across semantic relationships. The incorporation of Zorn's Lemma for handling infinite sets and the inherent ranking mechanism for finite subsets ensures both mathematical consistency and algorithmic feasibility, making this model particularly robust for analyzing complex, interconnected systems where traditional either fall short or are computationally inefficient. By representing entities as triplets ($v$ $\otimes$ $e$), the framework enables precise semantic reasoning in applications such as resource allocation and task scheduling.   

\subsection{The Core Theorem}
\label{sec:theorem}
Let $H = (V, E, \Omega)$ be a hyperstructure defined from the hypergraph $(V,E)$ along with a set of semantic operators $\Omega$ and let $T = V \times E \times \Omega$ be the set of semantic entities, then we have the following.
\begin{theorem}[Computational Semantic Ordering]
For any hyper-structure $H$, there exists a partial order $\leq$ on $T$ and a family of functors $F: T \rightarrow T$ such that:
\begin{enumerate}
\setlength{\itemsep}{0pt}
\item \textbf{(Ordering Property)} For any $\tau_1, \tau_2 \in T$, $\tau_1 \leq \tau_2$ if and only if:
\( \omega_1(v_1, e_1) \sqsubseteq \omega_2(v_2, e_2) 
\) where $\sqsubseteq$ is a semantic ordering induced by the operator.

\item \textbf{(DAG Property)} There exists a functor $G: T \rightarrow \mathcal{D}$ where $\mathcal{D}$ is the category of directed acyclic graphs such that:
\begin{itemize}
    \setlength{\itemsep}{0pt} 
    \item $G$ preserves the partial order structure
    \item The image $G(T)$ admits an efficient topological sorting
    \item The ordering induced by $G$ respects semantic weights
\end{itemize}
\item \textbf{(Completeness Property)} For any chain $C \subseteq T$:
\begin{itemize}
    \setlength{\itemsep}{0pt}
    \item $\sup(C)$ exists if any $\omega \in C$ is expansive (by Zorn's Lemma)
    \item $\inf(C)$ exists if any $\omega \in C$ is contractive (by Zorn's Lemma)
    \item Both operations are computationally realizable through categorical (co)limits
\end{itemize}
\item \textbf{(Functorial Property)} The family of functors $F$ satisfies:
\begin{itemize}
    \setlength{\itemsep}{0pt}
    \item Preserves both partial and induced total orders
    \item Respects operator semantics through categorical composition
    \item Maintains hyperedge consistency via limit preservation
    \item Admits efficient computational implementation
\end{itemize}

\end{enumerate}
\end{theorem}

Furthermore, this structure supports both the original ranking and compositional extensions while guaranteeing computational feasibility.

\begin{figure}
    \centering
    \includegraphics[width=\columnwidth]{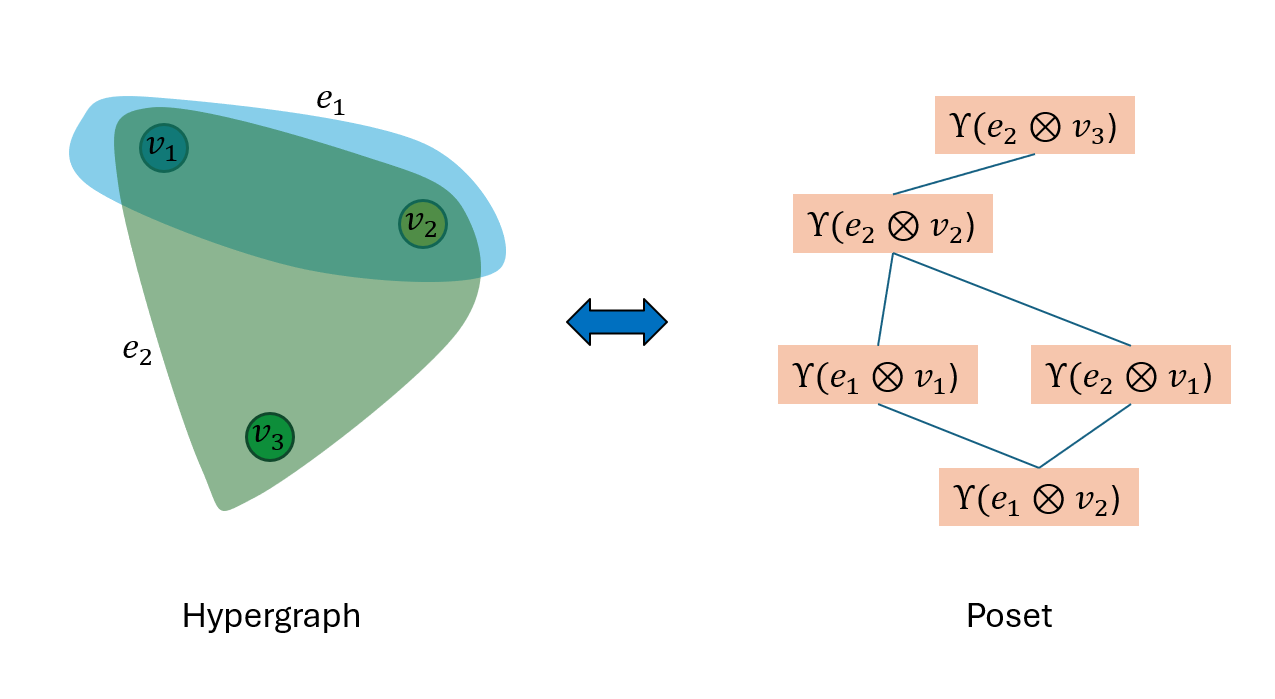}
    \caption{Existence of a Poset in Hypergraph}
    \label{fig:single1}
\end{figure}

This theorem establishes the existence of a computational semantic ordering within a hyperstructure, demonstrating how semantic relationships between entities can be rigorously structured and efficiently manipulated.  It asserts that for any hyperstructure, a partial order can be defined on the set of semantic entities, reflecting the semantic relationships induced by categorical limits in the form of a poset Fig.~\ref{fig:single1}. Furthermore, it guarantees the existence of functors that preserve this order and map the semantic entities to a category of directed acyclic graphs, enabling efficient topological sorting and respecting semantic weights.  Crucially, the theorem ensures the completeness of this ordering, the existence of suprema and infima for chains of semantic entities, and the functorial properties necessary for maintaining consistency and enabling efficient computation, supporting both original and compositional extensions of the ranking process.

\subsection{The Approximation Bound}~\label{sec:bound}
\begin{figure}
\centering
    \includegraphics[width=\columnwidth]{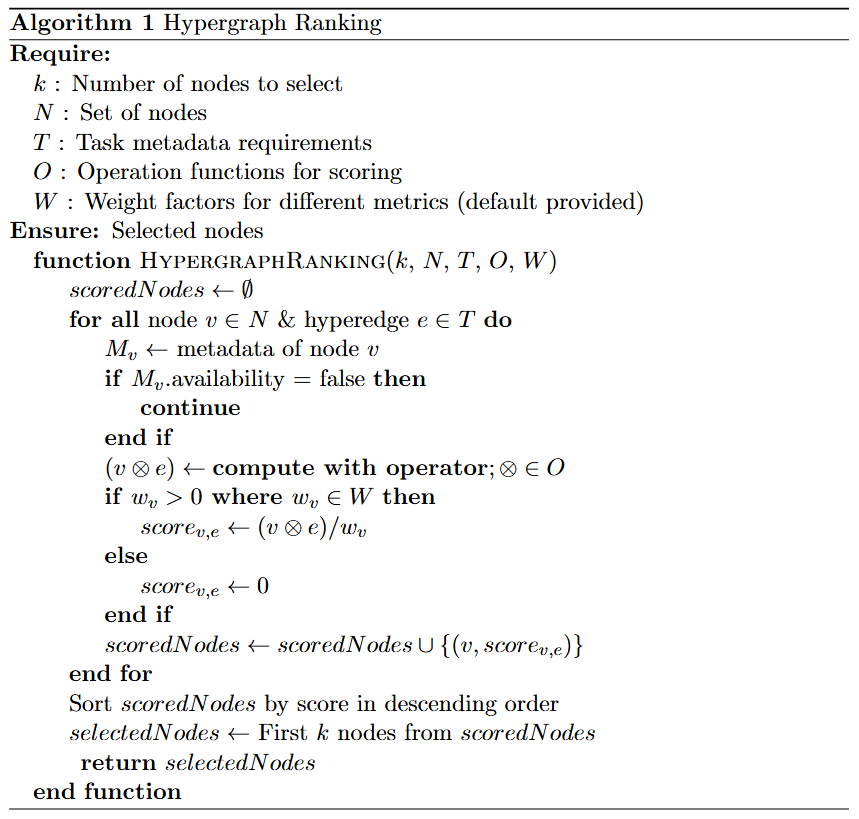}
    \caption{Hypergraph based Ranking Algorithm.}
    \label{fig:alg}
\end{figure}

The \textbf{approximation factor} for the algorithm in Fig.~\ref{fig:alg} depends on how well $e_{\text{alg}}$ (the selected top-$k$ resources for hyperedge $e$) approximates $e_{\text{opt}}$ (the optimal selection). Using semantic ordering reduces the misalignment cost. The approximation factor $\alpha$ is defined as:
\[
\frac{C_{\text{alg}}(e)}{C^*(e)} = \frac{\sum_{v \in e_{\text{alg}}} w(v)}{\sum_{v \in e_{\text{opt}}} w(v)}
\]
where ${C_{\text{alg}}(e)}$ is the cost of allocated resources by our algorithm and ${C^*(e)}$ is the optimal cost. This is expressed in terms of sum of the weights responsible for the cost of a resource. 
The relevance score $\Upsilon$ ($v$, $e$) of a resource $v$ in a task/hyperedge $e$ is defined as: 
$\Upsilon$(v, e) = ($v$ $\otimes$ $e$) / $w$($v$). The approximation factor $\alpha$ can be expressed as:

\[
\alpha = \frac{\sum_{v \in e_{\text{alg}}} w(v)}{\sum_{v \in e_{\text{opt}}} w(v)} = \frac{\sum_{v \in e_{\text{alg}}} (v \otimes e) / \Upsilon(v,e) }{\sum_{v \in e_{\text{opt}}} w(v)}
\]
From the expression above, we can observe that we need to bound \[\sum_{v \in e_{\text{alg}}} (v \otimes e) / \Upsilon(v,e).\]
For a fixed vertex \(v^*\) in the top-$k$ ranked entities, the value 
 \[
\frac{(v^* \otimes e)} {\Upsilon(v^*,e)} = \frac{1} {\Upsilon(v^*,e)} \sum_{i \in I}  \mu_i f_i(v^*,e) = \frac{1} {\Upsilon(v^*,e)} \sum_{i \in I}  \mu_i f_i(e)  
\]
where $i$ runs over the index set $I$ covering all the metadata variables of the vertex $v^*$, \(\mu_i\) are the weights for the metadata variable specific for ranking and \(f_i\) are functions of single variable. The function \(f_i(e)\) evaluates the semantic meaning of the $i-th$ metadata variable in the vertex $v^*$ and hyperedge $e$. Using Triangle Inequality on weighted sum of the functions \(f_i(e)\) , 
\[
|\sum_{i \in I}  \mu_i f_i(e) |  \leq \sum_{i \in I}  \mu_i |f_i(e) | = M
\]
By construction \(\Upsilon(v^*,e)\) will be a positive quantity greater than 1. For all the top-$k$ vertices in the hypergraph ranking, the cost is bounded as:
\[
\sum_{v \in e_{\text{alg}}} \frac{(v \otimes e)}{\Upsilon(v,e)} \leq k.M
\]
Therefore, the approximation factor \(\alpha \) satisfies:
\[
\alpha \leq k.M.\sum_{v \in e_{\text{opt}}} w(v) 
\]

\[
\alpha \leq k.M. C^*(e) 
\]

\section{Performance Analysis}\label{sec:perf}
\begin{figure*}[t]    
    \centering
    \includegraphics[width=\textwidth]{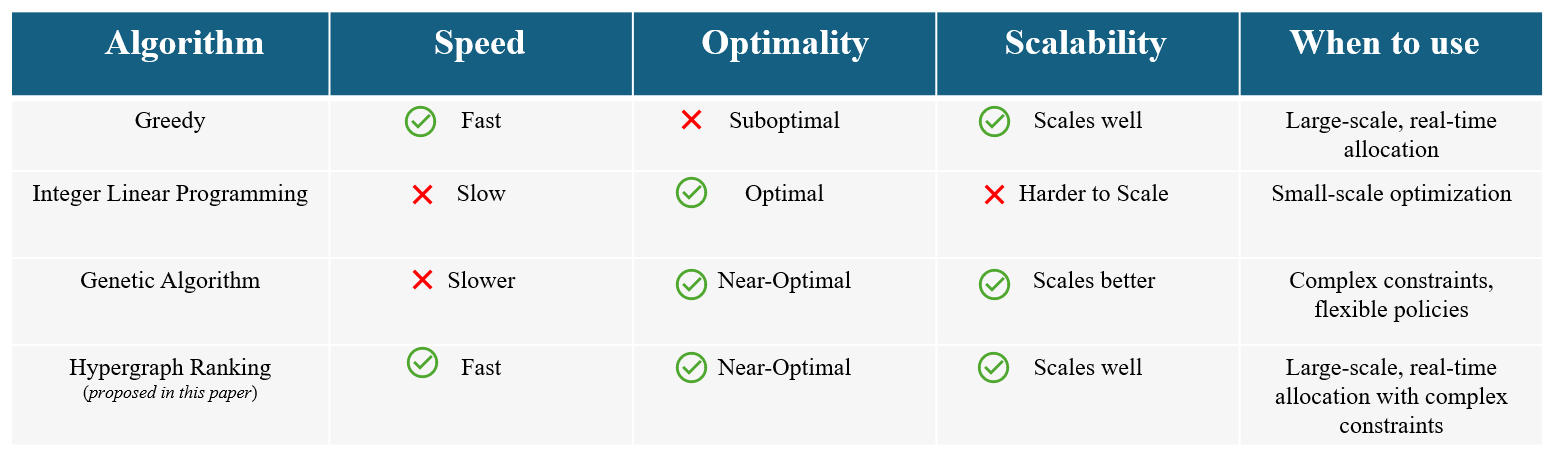}
    \caption{The comparison between various algorithms.}
    \label{fig:perf}
\end{figure*}
The core idea is that resources and tasks are not independent entities but interdependent in a structured manner. The algorithm models this as a hypergraph, where nodes represent resources and hyperedges represent tasks embedded with multi-way relationships between them. Instead of treating resources as independent entities (like in greedy approaches) or solving an optimization problem in a linear fashion (like ILP), the proposed method leverages semantic operators to ensure that only the most semantically relevant resources are assigned to tasks.
Greedy allocation \& myopic, does not optimize long-term allocation. The proposed ranking method performs ranked projections, meaning it finds globally optimal allocations across tasks. Integer Linear Programming (ILP) approach is computationally expensive and rigid. The hypergraph approach embeds interactions among the metadata and input variables, reducing computational complexity while maintaining better approximations. 
Heuristic-based methods like Genetic Algorithm (GA) have slow convergence and very difficult to model complex constraints. Arbitrary weight assignment methods lead to suboptimal allocations and inconsistent costs of allocation. In hypergraph ranking algorithm, the bounded operator ensures semantic alignment, leading to more meaningful rankings of resource-task pairs. The ranking-based projection method mathematically guarantees better approximation factors.\\

Instead of allocating resources greedily or solving a static optimization problem, the algorithm projects relevance scores onto the hypergraph structure. The bounded operator ($\otimes$) ensures that resource-task assignments are mathematically well-defined, avoiding extreme allocations. The algorithm achieves a tighter bound than naive weight-based approaches. This means it reduces the discrepancy between the optimal solution and the computed solution, providing near-optimal performance. Unlike greedy or heuristic-based methods, this algorithm ensures resources are semantically aligned with task needs rather than being assigned based on static weights. Unlike ILP, which becomes infeasible for large-scale problems, the hypergraph approach performs ranking projections efficiently, making it computationally feasible. The ranking function ensures that changes in task-resource interactions dynamically adjust allocations, unlike genetic algorithms, which may take a long time to converge. A comparison table is shown in Fig.~\ref{fig:perf} to highlight the key advantages/disadvantages of each of these methods. This approach bridges the gap between theory and practice, making resource allocation not just optimal but also explainable and interpretable.

\section{Applications \& Simulations}\label{sec:simulation}

In this section, we apply our framework from Section~\ref{sec:modeling} to resource allocation and task scheduling, emphasizing the performance gains of our approach compared to common approaches from the literature through simulations. Then, we describe the applicability of our approach to \emph{Text-to-SQL}~\cite{spider} problems. Finally, we discuss implementational considerations to improve the runtime of our solution for large numbers of metadata variables and node count. 

\begin{figure}
\centering
    \includegraphics[width=\columnwidth]{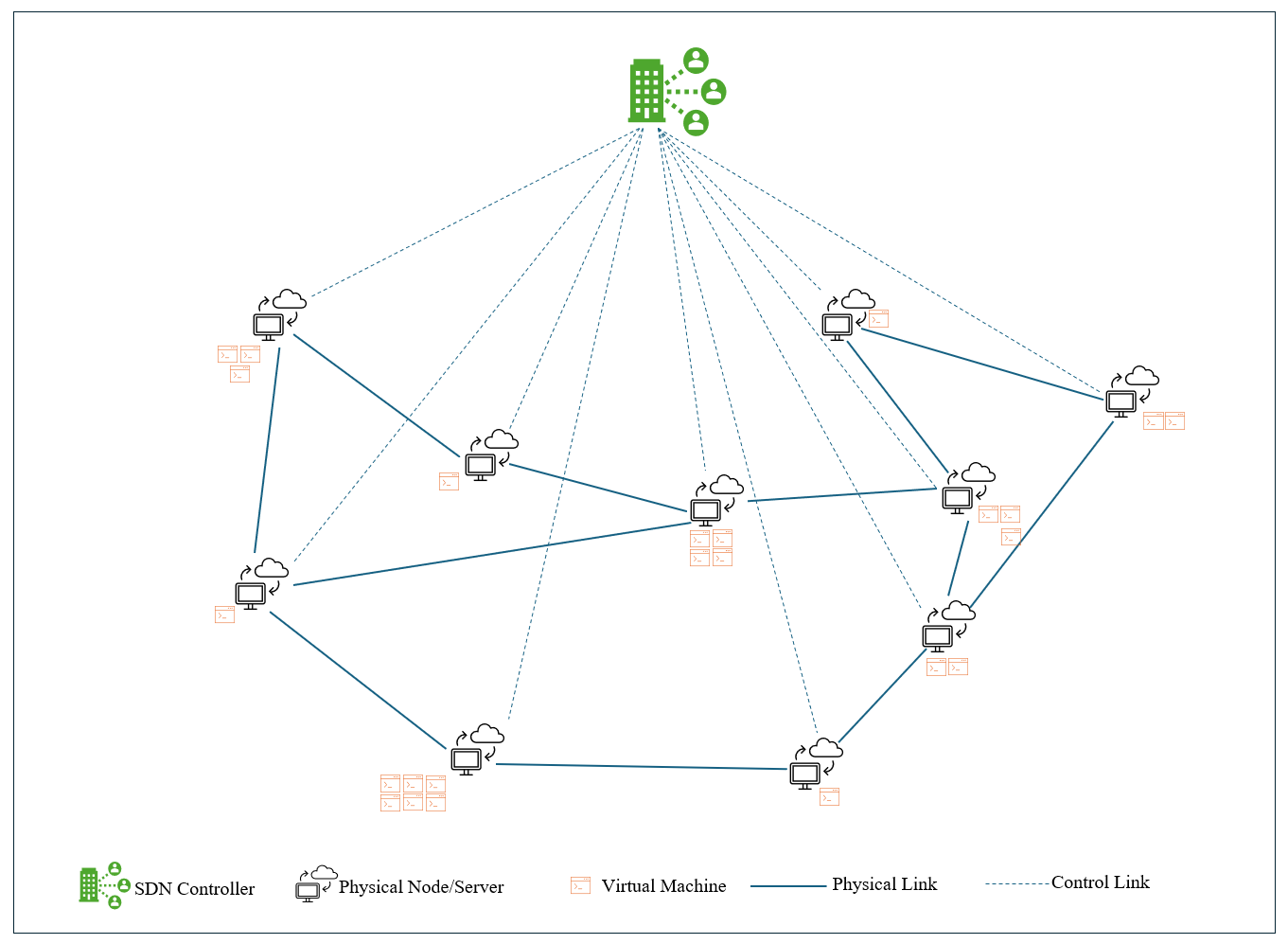}
    \caption{Illustration of the core domain in end-to-end network slicing}
    \label{fig:net_slice}
\end{figure}

\subsection{Resource Allocation}
In the basic resource allocation problem (network slicing), we are given a pool of resource and a task with its resource requirements, and the goal is to assign resources for to the task minimizing the over-utilization of resources. We depict an example core domain in network slicing scenario in Fig.~\ref{fig:net_slice}.  Mixed Integer Linear Programming~\cite{8349954} (ILP) and Random Allocation~\cite{LI201212213} are common solutions to resource allocation, where ILP yields an optimal solution with respect to the total cost of allocated resources while Random Allocation yields an approximate solution. Here we run simulations to compare two resource allocation strategies: our hypergraph ranking-base approach and ILP. The experimental setup involves generating a random graph with $N$ nodes, where each node is assigned metadata across six attributes: CPU cores, RAM, storage, bandwidth, latency, and cost. 

\begin{figure}
\centering
    \includegraphics[width=\columnwidth]{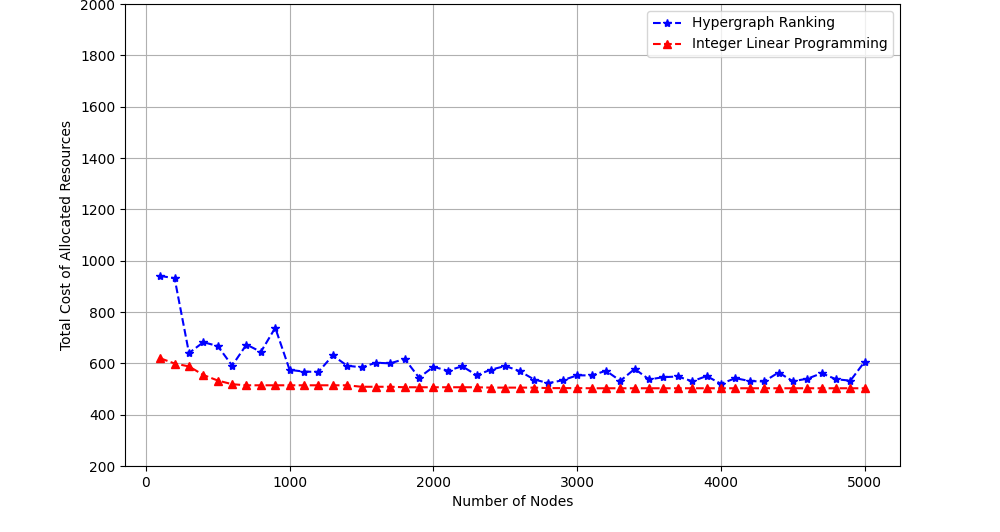}
    \caption{Total cost of allocated resources for Hypergraph Ranking and ILP.}
    \label{fig:ilp_vs_hr}
\end{figure}

\begin{figure}
\centering
    \includegraphics[width=\columnwidth]{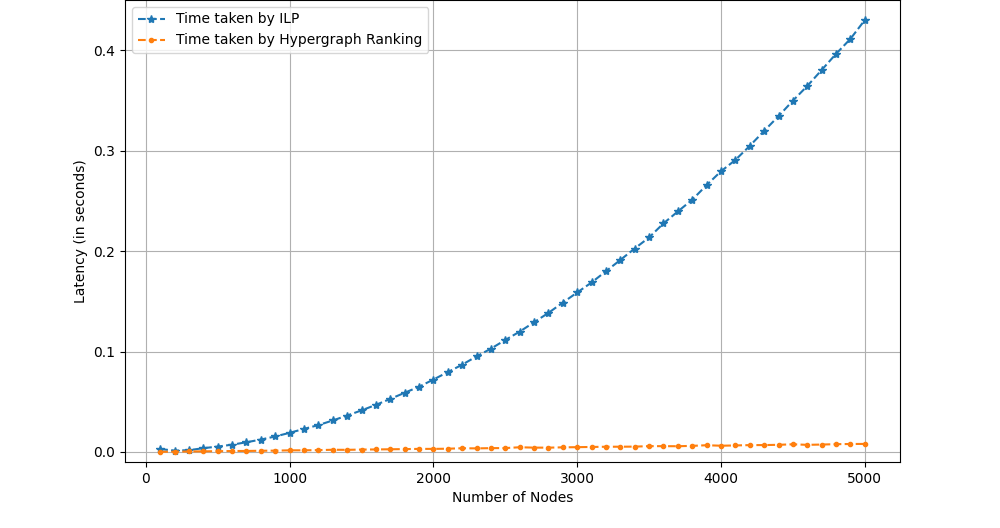}
    \caption{Latency of ILP versus Hypergraph Ranking-based approach (our).}
    \label{fig:latency_improvement}
\end{figure}

In the first set of experiments, a task with specific resource requirements is defined, and the allocation algorithms aim to select $k$ nodes (in this case, $k=5$) that minimize a composite score based on the operator semantics defined between task requirements and node capabilities. We vary the number of nodes from 100 to 5000, employing custom distance functions for each metadata type and tracking the total allocation cost and computational latency. In Fig.~\ref{fig:ilp_vs_hr}, we compare the total cost of allocated resources for ILP and our Hypergraph Ranking-approach, showing that our solution approaches the optimal results of ILP for the networks of practical sizes. The overhead of our approach is the price we pay for the multifold latency improvement depicted in Fig.~\ref{fig:latency_improvement}.

\begin{figure}
\centering
    \includegraphics[width=\columnwidth]{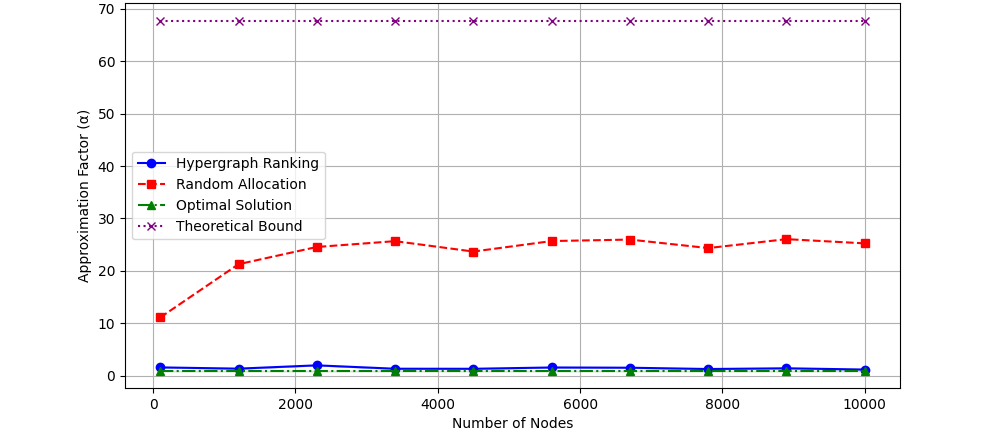}
    \caption{Quality of Approximation for Hypergraph Ranking-based approach (our) and Random Allocation. Theoretical bound from~Section~\ref{sec:bound}.}
    \label{fig:theory}
\end{figure}

In the second set of experiments, we compare the quality of approximation of our approach against Random Allocation. As shown in Fig.~\ref{fig:theory}, our approach offers far superior approximation compared to Random Allocation.

\subsection{Scheduling}
The application of our proposed algorithm can easily include scheduling problems. The system models tasks and computing nodes as a hypergraph, where nodes represent computing resources, and hyperedges represent tasks with specific resource demands. The algorithm dynamically evaluates node-task assignments using a scoring function that integrates multiple resource constraints, including CPU cores, RAM, execution time, cost, etc. The operator between nodes and hyperedges can be chosen in such a way that the priorities, conflict resolution, deadlocks, etc. are embedded in the ranking.
The relevance score $\Upsilon(v,e)$ to each node $v$ for a given task $e$ using the function:

\[
\Upsilon(v, e) = \frac{(v \otimes e)}{w(v)}
\]

where $(v \otimes e)$ denotes the priority/requirement of node for the task attributes, and $w(v)$ represents the total resource weight. The ranking process ensures that the most cost-effective and prioritized node is chosen, minimizing the total allocation cost while maintaining resource efficiency.

Traditional scheduling approaches such as First-Come-First-Serve (FCFS), Round Robin (RR), and Shortest Job First (SJF) are widely used in resource allocation. However, they have several limitations:

\begin{itemize}
    \setlength{\itemsep}{0pt} 
    \item \textbf{FCFS}: Assigns tasks in the order they arrive, leading to potential bottlenecks if earlier tasks require extensive resources.
    \item \textbf{Round Robin (RR)}: Allocates equal CPU time to each task but does not account for varying resource needs, leading to inefficiencies.
    \item \textbf{Shortest Job First (SJF)}: Prioritizes tasks with the shortest execution time but ignores resource constraints, potentially overloading certain nodes.
    \item \textbf{Integer Linear Programming (ILP)}: Provides optimal solutions but is computationally expensive for large-scale dynamic scheduling.
\end{itemize}

In contrast, the hypergraph ranking approach considers CPU, RAM, execution time, cost and other metadata variables simultaneously. It ensures that tasks are assigned to nodes that minimize cost while maximizing efficiency. Unlike ILP, which becomes computationally intractable with a large number of tasks, this approach maintains efficiency in large-scale scheduling. And it avoids assigning tasks to unavailable or overloaded nodes, improving system reliability.
The Hypergraph ranking algorithm is applicable in various real-world scheduling scenarios, including dynamic allocation of cloud instances to user workloads, optimizing resource utilization, scheduling tasks on supercomputers while minimizing execution time, assigning computational tasks to edge nodes based on their processing capability and efficiently managing virtual machines for distributed computing frameworks like Apache Spark or Kubernetes.

\begin{figure}
\centering
    \includegraphics[width=\columnwidth]{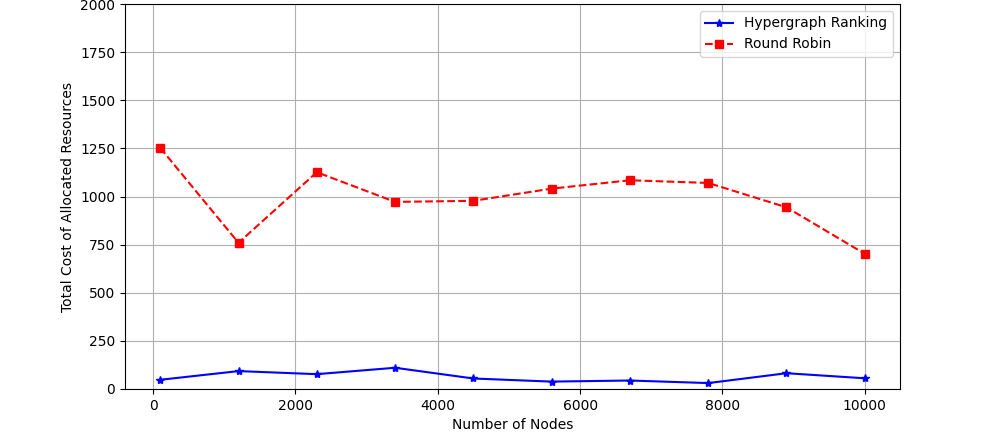}
    \caption{Round Robin versus Hypergraph Ranking}
    \label{fig:scheduling}
\end{figure}

Fig.~\ref{fig:scheduling} shows total cost of resources allocated for the scheduled task on a cloud-based job scheduling system where tasks with varying computational demands must be assigned to virtual machines (VMs). For the experimental setup, three tasks as specified below are considered:

\begin{itemize}
    \setlength{\itemsep}{0pt} 
    \item \textbf{Task 1}: Requires 8 CPU cores, 16GB RAM, and has an execution time of 5 seconds.
    \item \textbf{Task 2}: Requires 4 CPU cores, 8GB RAM, and has an execution time of 10 seconds.
    \item \textbf{Task 3}: Requires 16 CPU cores, 32GB RAM, and has an execution time of 2 seconds.
\end{itemize}

The cloud infrastructure consists of 500 VMs with varying configurations. The algorithm evaluates the resource difference functions for CPU, RAM, execution time, and cost, then ranks available VMs based on their suitability for each task. The best-fit VM is selected for each task, ensuring minimum resource wastage and optimal cost efficiency. The hypergraph ranking approach provides a scalable and cost-aware solution for dynamic resource scheduling in complex, heterogeneous environments. It outperforms traditional scheduling methods by incorporating multiple resource constraints, optimizing cost, and maintaining scalability. The results demonstrate that this approach provides near-optimal task assignments with significantly lower computational complexity than ILP or Round Robin, making it highly suitable for modern cloud and high-performance computing environments.
The implemented scheduling framework compares Hypergraph Ranking and Round Robin (RR) Scheduling in terms of total resource allocation cost across varying network sizes (100 to 500 nodes). The Hypergraph Ranking method optimizes task assignments by evaluating a weighted scoring function that considers multiple resource constraints (CPU, RAM, execution time, and cost). This approach has a computational complexity of $\mathcal{O}(n \log n)$ due to the sorting operation involved in ranking nodes based on their suitability scores. Conversely, the Round Robin scheduler assigns tasks sequentially in a cyclic manner, exhibiting a lower computational complexity of $\mathcal{O}(n)$ but at the cost of suboptimal resource allocation. Experimental results indicate that Hypergraph Ranking achieves a lower total cost by intelligently selecting the most cost-efficient nodes, whereas RR, despite its simplicity, incurs higher costs due to its lack of resource awareness. The trade-off between scheduling efficiency and computational overhead is particularly relevant for large-scale distributed computing environments, where intelligent scheduling can significantly reduce resource misallocation and operational expenses. The cost analysis, coupled with the complexity comparison, demonstrates the practical advantage of optimization-based task allocation over fairness-driven heuristics.

\subsection{Table Selection in Text-to-SQL}
\begin{figure}
\centering
    \includegraphics[width=\columnwidth]{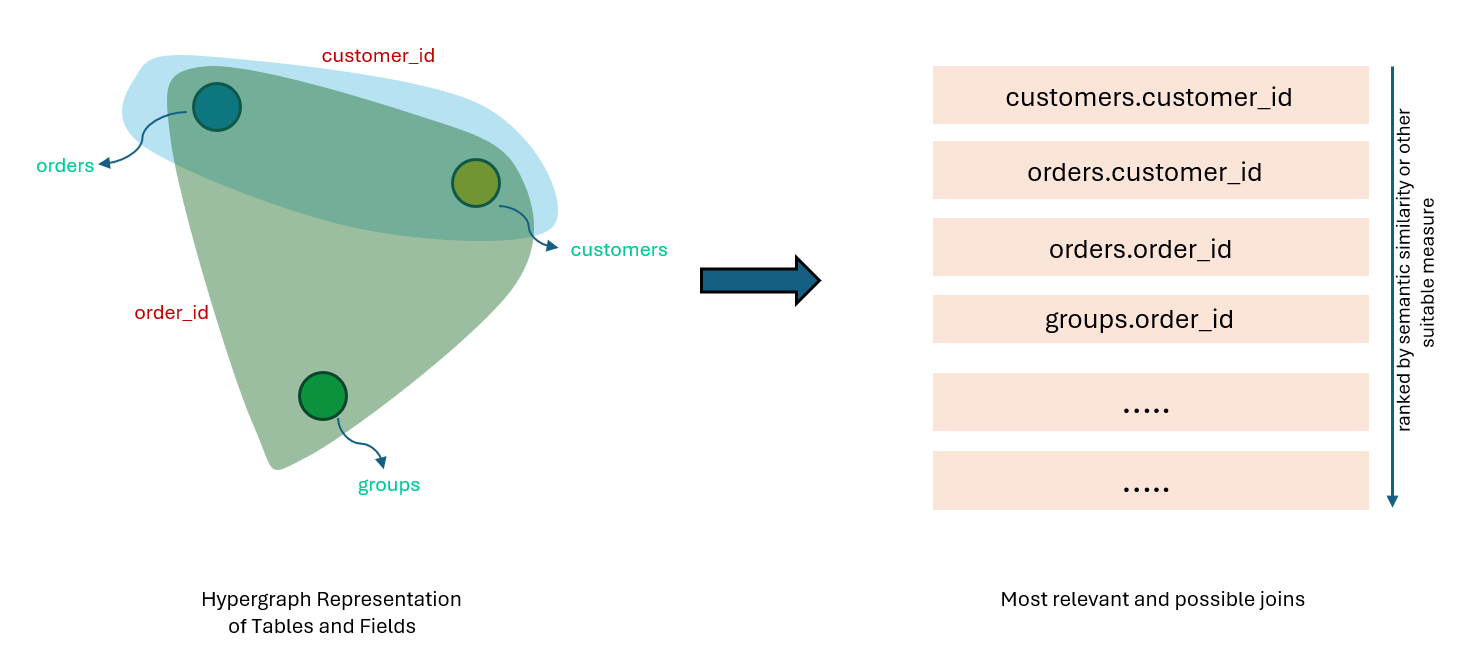}
    \caption{Illustration of hypergraph ranking in Text-to-SQL}
    \label{fig:table_selection}
\end{figure}

In Text-to-SQL, the retrieval of correct tables along with foreign-key relations need to be considered carefully when dealing with LLMs to give an accurate SQL query. The hypergraph representation and ranking can dynamically capture the most relevant relations among all the tables retrieved. This retrieval component uses a prior knowledge about the possible joins from the ground truth. When dealing with a new question, the possibility of joining two tables is highly dependent on the context and the database structure. For an enterprise scale database, it could lead to a large number of possible joins, whereas only a few are required to generate the right SQL query. As shown in the Fig.~\ref{fig:table_selection}, one hyperedge which is a “field” or “column” in this case, could be relating 5 tables or just one table depending on the schema and foreign keys. The hypergraph representation is suitable in this case, as the output which is required is in the form of "table.column". The table name is concatenated with the field name and semantic similarity is performed on this concatenated entity. The concatenation contains information about the table and field at the same time. So, retrieval gets most relevant “joins” in this dynamically constructed hypergraph representation. The entity \((v \otimes e)\) is the string concatenation of table names (\textit{nodes}) and the field names (\textit{hyperedges}). The ranking is performed among these entities on the basis of cosine similarity to input natural language question. This concatenation can be further enhanced by adding an additional natural language context to the entity \((v \otimes e)\) helping in better retrieval and ranking.

\subsection{Implementational Optimizations}
Resource allocation is a time sensitive task in cloud computing, software defined networking, and distributed AI inference/training, and naive implementations of our approach would scale poorly for large number of metadata variables and network size due to sequential execution bottlenecks. The use of template programming for compile-time function selection would help in scalability without runtime overhead. The approach for metadata matching discussed above with hypergraph ranking is inherently parallel and can be leveraged using C++20's execution policies. For five metadata variables crucial for allocation resources for AI workloads, i.e. CPU cores, RAM, GPU Memory, Disk Speed, and network bandwidth, the custom matching functions must be implemented to compare AI model requirements with cloud instance specifications:
\begin{lstlisting}[language=C++, caption=Metadata Matching Functions]
double custom_cpu_function(double node_value, double task_value) {
return std::abs(node_value - task_value);}

double custom_ram_function(double node_value, double task_value) {...}

double custom_gpu_function(double node_value, double task_value) {...}

double custom_disk_function(double node_value, double task_value) {...}

double custom_network_function(double node_value, double task_value) {...}
\end{lstlisting}
C++ tuples can then be used for function selection at compile-time, avoiding runtime overhead (see Appendix~\ref{appendix:impl} for details). In addition, parallel execution can significantly improves the performance of the implementation. The following code snippet demonstrates metadata matching in parallel using C++ \texttt{std::execution::par}. With this parallelized version, the algorithm will speedup approximately five times due to simultaneous execution of the five metadata functions.
\begin{lstlisting}[language=c++, caption=Parallel Computation of Scores across multiple nodes]
std::transform(std::execution::par, 
nodes.begin(), nodes.end(), scores.begin(), [&](const Node& node) { return compute_total_score(node, task_metadata); });
\end{lstlisting}

\section{Related Works}\label{sec:related}
Researchers have proposed deep reinforcement learning (DRL) based solutions to networking slicing\cite{9846944} \cite{8943940} \cite{Zheng2024EfficientRA} \cite{li2016endtoendnetworkslicingframework}. These approaches require a large number of online interactions with the real networking systems and deployment of such solutions faces several system level challenges including preparation of networking data and providing appropriate network programmability for DRL. These solutions struggle to scale up to large networks as well as large number of metadata variables. While a hierarchical approach \cite{article} tries to solve the mutually coupled inter-slice and intra-slice problem in allocating radio and computing resources, it still suffers from the probabilistic decision making for size and number of resource blocks. There are very few graph based\cite{GUPTA2025103029} formulations of the resource allocation problem and its automation\cite{PERDIGAO2025107991}. The dynamic nature of resource allocation \cite{sidali2023onlineoptimizationnetworkresource} in virtual machine migration 
 optimization\cite{gong2024dynamicresourceallocationvirtual} poses even a harder challenge as it updates the metadata for all available resources in real time. The major bottleneck to handle in these cases is the speed of inference. Robustness is another challenge in network slicing where certainty of number of users requests, amount of bandwidth and requested virtual network functions workloads is not guaranteed due to the dynamic and unpredictable nature of network traffic patterns, user mobility, and varying service demands. This uncertainty\cite{unknown} necessitates adaptive resource allocation mechanisms that can handle fluctuations while maintaining Quality of Service (QoS) requirements and Service Level Agreements (SLAs). There exists measurements based methods\cite{popescu2024measurementbasedresourceallocationcontrol} for resource allocation and control in data centers. The network monitoring data from these setups can be directly served as an input to the proposed hypergraph ranking algorithm for resource allocation to get better results. The mixed integer programming formulations\cite{8349954} of resource allocation show promising accuracy but are not easily extensible to complex constraints. A critical challenge in modern mobile networks is to efficiently managing the coexistence of traditional real-time traffic with delay-tolerant IoT communications. The authors \cite{10.5555/3504035.3504129} propose an innovative reinforcement learning approach that dynamically schedules IoT traffic while preserving quality of service for conventional applications, demonstrating a significant 14.7\% increase in network capacity through real-world testing in Melbourne's downtown area.
Retrieval of right tables and their foreign key relations in databases is a key challenge in Text-to-SQL domain. The state of the art methods utilize integer linear programming \cite{chen2024tableretrievalsolvedproblem} or AL/ML based agentic workflows\cite{wang2021ratsqlrelationawareschemaencoding} 
\cite{lee2024hybgraghybridretrievalaugmentedgeneration} which lack the right knowledge representation of the database structure.  
There is very little hypergraph ranking work proposed in the literature. Ranking of researchers' skills via hypergraphs
\cite{DBLP:journals/peerjpre/KongLYYBX19} was proposed but the mathematical richness of their ranking model and scalability aspects were not discussed.

\section{Conclusion}\label{sec:conclusion}
We introduced a novel class of hypergraph ranking algorithms, representing a significant advancement in resource allocation and scheduling methodologies and transcending traditional approaches like integer linear programming (ILP) and stochastic approximation algorithms. By projecting relevance scores onto the hypergraph structure and utilizing a bounded operator, the algorithm achieves near-optimal resource-task assignments with superior computational efficiency and semantic alignment. Unlike greedy heuristics or static optimization methods, this approach dynamically adapts to changing task-resource interactions, providing a scalable solution that maintains computational feasibility even for large-scale networks. Our algorithmic framework introduced a novel paradigm that bridges theoretical optimization principles with practical implementation, offering not just computational efficiency but also interpretability. Compared to the slow convergence of stochastic approximation approaches and ILP's computational intractability, the hypergraph ranking method emerges as a versitile, adaptive strategy for distributed resource allocation, task scheduling, and other problems that fit the hypergraph modeling paradigm. The novel framework we proposed opens a vast exploration space for more complex semantic alignment techniques that will further reduce allocation discrepancies at negligible computational cost.

\bibliography{references}
\bibliographystyle{plain}

\section{Appendix}\label{appendix:impl}
\begin{lstlisting}[language=c++, caption=Parallel Hypergraph Ranking Implementation]
#include <tuple>
#include <functional>
#include <vector>
#include <execution>  // Parallel execution
#include <iostream>

using MetadataFunctionTuple = std::tuple<
    std::function<double(double, double)>,  // CPU
    std::function<double(double, double)>,  // RAM
    std::function<double(double, double)>,  // Storage
    std::function<double(double, double)>,  // Bandwidth
    std::function<double(double, double)>   // Latency
>;

MetadataFunctionTuple metadata_functions = {
    custom_cpu_function,
    custom_ram_function,
    custom_storage_function,
    custom_bandwidth_function,
    custom_latency_function
};

template<size_t Index, typename... Functions>
double execute_function(size_t metadata_index, double node_value, double task_value, std::tuple<Functions...>& functions) {
    if constexpr (Index < sizeof...(Functions)) {
        if (metadata_index == Index) {
            return std::get<Index>(functions)(node_value, task_value);
        }
        return execute_function<Index + 1>(metadata_index, node_value, task_value, functions);
    }
    return 0.0;
}

double compute_metadata(size_t metadata_index, double node_value, double task_value) {
    return execute_function<0>(metadata_index, node_value, task_value, metadata_functions);
}


double custom_cpu_function(double node_value, double task_value) {
    return std::min(node_value, task_value) / std::max(node_value, task_value);
}

double custom_ram_function(double node_value, double task_value) {
    return (node_value >= task_value) ? 1.0 : node_value / task_value;
}

double custom_storage_function(double node_value, double task_value) {
    return std::log(1 + std::min(node_value, task_value)) / std::log(1 + std::max(node_value, task_value));
}

double custom_bandwidth_function(double node_value, double task_value) {
    return node_value / (task_value + 1);
}

double custom_latency_function(double node_value, double task_value) {
    return 1.0 / (1.0 + node_value / task_value);
}
struct Node {
    int id;
    std::vector<double> metadata_values;  // CPU, RAM, Storage, Bandwidth, Latency
};

// Compute total metadata relevance score
double compute_total_score(const Node& node, const std::vector<double>& task_metadata) {
    double total_score = 0.0;
    for (size_t i = 0; i < node.metadata_values.size(); ++i) {
        total_score += compute_metadata(i, node.metadata_values[i], task_metadata[i]);
    }
    return total_score;
}

int main() {

//Sample metadata about nodes - Change this for your application
    std::vector<Node> nodes = {
        {1, {16, 32, 2.0, 500, 10}}, {2, {8, 16, 1.0, 300, 20}},
        {3, {32, 64, 4.0, 800, 5}}, {4, {4, 8, 0.5, 150, 30}},
        {5, {12, 24, 1.5, 400, 15}}, {6, {64, 128, 8.0, 1000, 2}}
    };

    std::vector<double> task_metadata = {16, 32, 2.0, 500, 10};

    std::vector<double> scores(nodes.size());

    std::transform(std::execution::par, nodes.begin(), nodes.end(), scores.begin(),
                   [&](const Node& node) { return compute_total_score(node, task_metadata); });

    std::cout << "Node Scores:\n";
    for (size_t i = 0; i < nodes.size(); ++i) {
        std::cout << "Node " << nodes[i].id << " -> Score: " << scores[i] << "\n";
    }

    return 0;
}


\end{lstlisting}

\section{Appendix}
\begin{lstlisting}
#include <iostream>
#include <vector>
#include <string>
#include <algorithm>
#include <map>
#include <functional>
#include <ranges>
#include <concepts>
#include <utility>

// Define Vertex and Hyperedge
typedef std::string Vertex;
typedef std::vector<Vertex> Hyperedge;

// Subset concept for compile-time validation of inputs
bool isSubset(const Hyperedge& subset, const Hyperedge& superset) {
    return std::ranges::all_of(subset, [&](const Vertex& elem) {
        return std::ranges::find(superset, elem) != superset.end();
    });
}

// Define SemanticEntity using modern C++ concepts
struct SemanticEntity {
    Vertex vertex;
  fff  Hyperedge edge;
    std::function<Hyperedge(const Hyperedge&)> operatorFn;

    SemanticEntity(Vertex v, Hyperedge e, std::function<Hyperedge(const Hyperedge&)> op)
        : vertex(std::move(v)), edge(std::move(e)), operatorFn(std::move(op)) {}

    Hyperedge applyOperator() const {
        return operatorFn(edge);
    }
};

// Partial Order Concept
struct PartialOrder {
    virtual bool compare(const SemanticEntity& e1, const SemanticEntity& e2) const = 0;
};

// PartialOrder Implementation
struct SemanticEntityOrder : public PartialOrder {
    bool compare(const SemanticEntity& e1, const SemanticEntity& e2) const override {
        return isSubset(e1.applyOperator(), e2.applyOperator());
    }
};

// Lattice using modern functional programming tools
struct Lattice {
    std::function<Hyperedge(const Hyperedge&)> meet;
    std::function<Hyperedge(const Hyperedge&)> join;

    Lattice(std::function<Hyperedge(const Hyperedge&)> m, std::function<Hyperedge(const Hyperedge&)> j)
        : meet(std::move(m)), join(std::move(j)) {}
};

// Ranking Function with improved low-latency data structures
std::map<SemanticEntity, double, std::function<bool(const SemanticEntity&, const SemanticEntity&)>> rankEntities(
    const std::vector<SemanticEntity>& entities) {

    auto comparator = [](const SemanticEntity& a, const SemanticEntity& b) {
        return a.applyOperator().size() < b.applyOperator().size();
    };

    std::map<SemanticEntity, double, decltype(comparator)> ranking(comparator);

    // Rank entities based on operator-applied edge size
    for (size_t i = 0; i < entities.size(); ++i) {
        ranking[entities[i]] = i + 1.0;
    }

    return ranking;
}

// Overload for std::map key comparison (SemanticEntity)
bool operator<(const SemanticEntity& lhs, const SemanticEntity& rhs) {
    return lhs.vertex < rhs.vertex || (lhs.vertex == rhs.vertex && lhs.edge < rhs.edge);
}

// Composition with modern functional concepts
SemanticEntity compose(const SemanticEntity& e1, const SemanticEntity& e2) {
    return SemanticEntity(
        e1.vertex + e2.vertex,
        e1.edge,  // Retain the edge of the first entity for simplicity
        [=](const Hyperedge& x) {
            return e2.operatorFn(e1.operatorFn(x));
        });
}

// Main Program for Testing
int main() {
    // Define example vertices, edges, and operators
    std::vector<Vertex> vertices = {"v1", "v2", "v3"};
    std::vector<Hyperedge> edges = {{"v1", "v2"}, {"v2", "v3"}, {"v1", "v3"}};
    std::vector<std::function<Hyperedge(const Hyperedge&)>> operators = {
        [](const Hyperedge& edge) { return Hyperedge(edge.rbegin(), edge.rend()); },
        [](const Hyperedge& edge) { return edge; },
        [](const Hyperedge& edge) { Hyperedge sorted = edge; std::ranges::sort(sorted); return sorted; }
    };

    // Create Semantic Entities
    std::vector<SemanticEntity> entities;
    for (size_t i = 0; i < vertices.size(); ++i) {
        entities.emplace_back(vertices[i], edges[i % edges.size()], operators[i % operators.size()]);
    }

    // Rank entities with low-latency data structures
    auto rankings = rankEntities(entities);
    for (const auto& [entity, rank] : rankings) {
        std::cout << "Vertex: " << entity.vertex << ", Rank: " << rank << "\n";
    }

    // Test composition with minimal overhead
    auto composedEntity = compose(entities[0], entities[1]);
    std::cout << "Composed Vertex: " << composedEntity.vertex << "\n";

    return 0;
}

\end{lstlisting}

\end{document}